\input harvmac


\def\lesssim{\mathrel{\mathpalette\fun <}}
\def\gtrsim{\mathrel{\mathpalette\fun >}}
\def\fun#1#2{\lower3.6pt\vbox{\baselineskip0pt\lineskip.9pt
  \ialign{$\mathsurround=0pt#1\hfil##\hfil$\crcr#2\crcr\sim\crcr}}}
\relax

\def\inbar{\,\vrule height1.6ex width.4pt depth0pt}
\font\cmss=cmss10 \font\cmsss=cmss10 at 7pt

\def\IC{\relax\hbox{$\inbar\kern-.3em{\rm C}$}}
\def\IR{\relax{\rm I\kern-.18em R}}
\def\IP{\relax{\rm I\kern-.18em P}}
\def\Z{\relax\ifmmode\mathchoice
{\hbox{\cmss Z\kern-.4em Z}}{\hbox{\cmss Z\kern-.4em Z}}
{\lower.9pt\hbox{\cmsss Z\kern-.4em Z}} {\lower1.2pt\hbox{\cmsss
Z\kern-.4em Z}}\else{\cmss Z\kern-.4em Z}\fi}

\lref\Ganor{
  O.~J.~Ganor,
  ``A note on zeroes of superpotentials in F-theory,''
  Nucl.\ Phys.\  B {\bf 499}, 55 (1997)
  [arXiv:hep-th/9612077].
}

\lref\Doran{
  C.~Doran, M.~Headrick, C.~P.~Herzog, J.~Kantor and T.~Wiseman,
  ``Numerical Kaehler-Einstein metric on the third del Pezzo,''
  arXiv:hep-th/0703057.
}

\lref\GR{
  B.~R.~Greene, C.~I.~Lazaroiu and M.~Raugas,
  ``D-branes on nonabelian threefold quotient singularities,''
  Nucl.\ Phys.\  B {\bf 553}, 711 (1999)
  [arXiv:hep-th/9811201].
}

\lref\DM{
  M.~R.~Douglas, B.~R.~Greene and D.~R.~Morrison,
  ``Orbifold resolution by D-branes,''
  Nucl.\ Phys.\  B {\bf 506}, 84 (1997)
  [arXiv:hep-th/9704151].
}

\lref\wijnholt{M. Wijnholt, ``Geometry of Particle Physics,''
arXiv:hep-th/0703047.}

\lref\louisone{
  M.~Grana, T.~W.~Grimm, H.~Jockers and J.~Louis,
  ``Soft supersymmetry breaking in Calabi-Yau orientifolds with D-branes  and
  fluxes,''
  Nucl.\ Phys.\  B {\bf 690}, 21 (2004)
  [arXiv:hep-th/0312232].
}

\lref\mirage{
  K.~Choi, A.~Falkowski, H.~P.~Nilles and M.~Olechowski,
  ``Soft supersymmetry breaking in KKLT flux compactification,''
  Nucl.\ Phys.\  B {\bf 718}, 113 (2005)
  [arXiv:hep-th/0503216].
}

\lref\DeWolfe{
  O.~DeWolfe and S.~B.~Giddings,
  ``Scales and hierarchies in warped compactifications and brane worlds,''
  Phys.\ Rev.\  D {\bf 67}, 066008 (2003)
  [arXiv:hep-th/0208123].
}

\lref\Maldacena{
  J.~M.~Maldacena,
  ``The large N limit of superconformal field theories and supergravity,''
  Adv.\ Theor.\ Math.\ Phys.\  {\bf 2}, 231 (1998)
  [Int.\ J.\ Theor.\ Phys.\  {\bf 38}, 1113 (1999)]
  [arXiv:hep-th/9711200].
}

\lref\OtherAdSCFT{
  S.~S.~Gubser, I.~R.~Klebanov and A.~M.~Polyakov,
  Phys.\ Lett.\  B {\bf 428}, 105 (1998)
  [arXiv:hep-th/9802109]\semi
  E.~Witten,
  ``Anti-de Sitter space and holography,''
  Adv.\ Theor.\ Math.\ Phys.\  {\bf 2}, 253 (1998)
  [arXiv:hep-th/9802150].
}

\lref\louistwo{
  H.~Jockers and J.~Louis,
  ``The effective action of D7-branes in N = 1 Calabi-Yau orientifolds,''
  Nucl.\ Phys.\  B {\bf 705}, 167 (2005)
  [arXiv:hep-th/0409098].
}

\lref\Berg{
  M.~Berg, M.~Haack and B.~Kors,
  ``String loop corrections to Kaehler potentials in orientifolds,''
  JHEP {\bf 0511}, 030 (2005)
  [arXiv:hep-th/0508043]\semi
  M.~Berg, M.~Haack and B.~Kors,
  ``On volume stabilization by quantum corrections,''
  Phys.\ Rev.\ Lett.\  {\bf 96}, 021601 (2006)
  [arXiv:hep-th/0508171].
}

\lref\Hebecker{
  A.~Hebecker and J.~March-Russell,
  ``The ubiquitous throat,''
  arXiv:hep-th/0607120.
}

\lref\GM{
  S.~B.~Giddings and A.~Maharana,
  ``Dynamics of warped compactifications and the shape of the warped
  landscape,''
  Phys.\ Rev.\  D {\bf 73}, 126003 (2006)
  [arXiv:hep-th/0507158].
}

\lref\ChoiRecent{
  K.~Choi and K.~S.~Jeong,
  ``Supersymmetry breaking and moduli stabilization with anomalous U(1) gauge
  symmetry,''
  JHEP {\bf 0608}, 007 (2006)
  [arXiv:hep-th/0605108].
}

\lref\RS{
  L.~Randall and R.~Sundrum,
  ``Out of this world supersymmetry breaking,''
  Nucl.\ Phys.\  B {\bf 557}, 79 (1999)
  [arXiv:hep-th/9810155].
}

\lref\other{
  G.~F.~Giudice, M.~A.~Luty, H.~Murayama and R.~Rattazzi,
  ``Gaugino mass without singlets,''
  JHEP {\bf 9812}, 027 (1998)
  [arXiv:hep-ph/9810442].
}

\lref\Ls{
  M.~A.~Luty and R.~Sundrum,
  ``Hierarchy stabilization in warped supersymmetry,''
  Phys.\ Rev.\  D {\bf 64}, 065012 (2001)
  [arXiv:hep-th/0012158].
}

\lref\Pomarol{
  A.~Pomarol and R.~Rattazzi,
  ``Sparticle masses from the superconformal anomaly,''
  JHEP {\bf 9905}, 013 (1999)
  [arXiv:hep-ph/9903448];
Z.~Chacko, M.~A.~Luty, I.~Maksymyk and E.~Ponton,
  ``Realistic anomaly-mediated supersymmetry breaking,''
  JHEP {\bf 0004}, 001 (2000)
  [arXiv:hep-ph/9905390]; E.~Katz, Y.~Shadmi and Y.~Shirman,
   ``Heavy thresholds, slepton masses and the mu term in anomaly mediated
  supersymmetry breaking,''
  JHEP {\bf 9908}, 015 (1999)
  [arXiv:hep-ph/9906296];
  K.~I.~Izawa, Y.~Nomura and T.~Yanagida,
  ``Cosmological constants as messenger between branes,''
  Prog.\ Theor.\ Phys.\  {\bf 102}, 1181 (1999)
  [arXiv:hep-ph/9908240];
  M.~Carena, K.~Huitu and T.~Kobayashi,
   ``RG-invariant sum rule in a generalization of anomaly mediated SUSY
  breaking models,''
  Nucl.\ Phys.\  B {\bf 592}, 164 (2001)
  [arXiv:hep-ph/0003187];
  B.~C.~Allanach and A.~Dedes,
  ``R-parity violating anomaly mediated supersymmetry breaking,''
  JHEP {\bf 0006}, 017 (2000)
  [arXiv:hep-ph/0003222];
I.~Jack and D.~R.~T.~Jones,
  ``R-symmetry, Yukawa textures and anomaly mediated supersymmetry  breaking,''
  Phys.\ Lett.\  B {\bf 491}, 151 (2000)
  [arXiv:hep-ph/0006116];
  D.~E.~Kaplan and G.~D.~Kribs,
  ``Gaugino-assisted anomaly mediation,''
  JHEP {\bf 0009}, 048 (2000)
  [arXiv:hep-ph/0009195];
  N.~Arkani-Hamed, D.~E.~Kaplan, H.~Murayama and Y.~Nomura,
  ``Viable ultraviolet-insensitive supersymmetry breaking,'' JHEP {\bf 0102}, 041 (2001) [arXiv:hep-ph/0012103];
  Z.~Chacko and M.~A.~Luty, 
  ``Realistic anomaly mediation with bulk gauge fields,''
  JHEP {\bf 0205}, 047 (2002)
  [arXiv:hep-ph/0112172];
  A.~E.~Nelson and N.~J.~Weiner,
   ``Gauge/anomaly Syzygy and generalized brane world models of  supersymmetry
  breaking,''
  Phys.\ Rev.\ Lett.\  {\bf 88}, 231802 (2002)
  [arXiv:hep-ph/0112210].}

\lref\Nelson{
  A.~E.~Nelson and M.~J.~Strassler,
  ``Exact results for supersymmetric renormalization and the supersymmetric
  flavor problem,''
  JHEP {\bf 0207}, 021 (2002)
  [arXiv:hep-ph/0104051]\semi
  A.~E.~Nelson and M.~J.~Strassler,
  ``Suppressing flavor anarchy,''
  JHEP {\bf 0009}, 030 (2000)
  [arXiv:hep-ph/0006251].}

\lref\Becker{
  K.~Becker and M.~Becker,
  ``M-Theory on Eight-Manifolds,''
  Nucl.\ Phys.\  B {\bf 477}, 155 (1996)
  [arXiv:hep-th/9605053].
}

\lref\GVW{
  S.~Gukov, C.~Vafa and E.~Witten,
  ``CFT's from Calabi-Yau four-folds,''
  Nucl.\ Phys.\  B {\bf 584}, 69 (2000)
  [Erratum-ibid.\  B {\bf 608}, 477 (2001)]
  [arXiv:hep-th/9906070].
}

\lref\DRS{
  K.~Dasgupta, G.~Rajesh and S.~Sethi,
  ``M theory, orientifolds and G-flux,''
  JHEP {\bf 9908}, 023 (1999)
  [arXiv:hep-th/9908088].
}

\lref\Herman{
  H.~L.~Verlinde,
  ``Holography and compactification,''
  Nucl.\ Phys.\  B {\bf 580}, 264 (2000)
  [arXiv:hep-th/9906182].
}

\lref\KMS{
  S.~Kachru, J.~McGreevy and P.~Svrcek,
  ``Bounds on masses of bulk fields in string compactifications,''
  JHEP {\bf 0604}, 023 (2006)
  [arXiv:hep-th/0601111].
}

\lref\Quevedo{
  P.~Candelas, A.~Font, S.~H.~Katz and D.~R.~Morrison,
  ``Mirror symmetry for two parameter models. 2,''
  Nucl.\ Phys.\  B {\bf 429}, 626 (1994)
  [arXiv:hep-th/9403187]\semi
  F.~Denef, M.~R.~Douglas and B.~Florea,
  ``Building a better racetrack,''
  JHEP {\bf 0406}, 034 (2004)
  [arXiv:hep-th/0404257]\semi
 V.~Balasubramanian, P.~Berglund, J.~P.~Conlon and F.~Quevedo,
  ``Systematics of moduli stabilisation in Calabi-Yau flux
  compactifications,''
  JHEP {\bf 0503}, 007 (2005)
  [arXiv:hep-th/0502058].
}

\lref\SS{
  M.~Schmaltz and R.~Sundrum,
  ``Conformal sequestering simplified,''
  JHEP {\bf 0611}, 011 (2006)
  [arXiv:hep-th/0608051].
}

\lref\GKP{
  S.~B.~Giddings, S.~Kachru and J.~Polchinski,
  ``Hierarchies from fluxes in string compactifications,''
  Phys.\ Rev.\  D {\bf 66}, 106006 (2002)
  [arXiv:hep-th/0105097].
}

\lref\RSone{
  L.~Randall and R.~Sundrum,
  ``A large mass hierarchy from a small extra dimension,''
  Phys.\ Rev.\ Lett.\  {\bf 83}, 3370 (1999)
  [arXiv:hep-ph/9905221].
}

\lref\students{W. Chuang and M. Mulligan, work in progress.}

\lref\KKLT{
  S.~Kachru, R.~Kallosh, A.~Linde and S.~P.~Trivedi,
  ``de Sitter vacua in string theory,''
  Phys.\ Rev.\  D {\bf 68}, 046005 (2003)
  [arXiv:hep-th/0301240].
}

\lref\KPV{
  S.~Kachru, J.~Pearson and H.~L.~Verlinde,
  ``Brane/flux annihilation and the string dual of a non-supersymmetric  field
  theory,''
  JHEP {\bf 0206}, 021 (2002)
  [arXiv:hep-th/0112197].
}

\lref\LS{
  M.~A.~Luty and R.~Sundrum,
  ``Radius stabilization and anomaly-mediated supersymmetry breaking,''
  Phys.\ Rev.\  D {\bf 62}, 035008 (2000)
  [arXiv:hep-th/9910202].
}

\lref\DFG{
  M.~Dine, P.~J.~Fox, E.~Gorbatov, Y.~Shadmi, Y.~Shirman and S.~D.~Thomas,
  ``Visible effects of the hidden sector,''
  Phys.\ Rev.\  D {\bf 70}, 045023 (2004)
  [arXiv:hep-ph/0405159].
}

\lref\Anisimov{A.~Anisimov, M.~Dine, M.~Graesser and S.~D.~Thomas,
  ``Brane world SUSY breaking,''
  Phys.\ Rev.\  D {\bf 65}, 105011 (2002)
  [arXiv:hep-th/0111235]\semi
  A.~Anisimov, M.~Dine, M.~Graesser and S.~D.~Thomas,
  ``Brane world SUSY breaking from string/M theory,''
  JHEP {\bf 0203}, 036 (2002)
  [arXiv:hep-th/0201256].
}

\lref\LSconf{
  M.~A.~Luty and R.~Sundrum,
  ``Supersymmetry breaking and composite extra dimensions,''
  Phys.\ Rev.\  D {\bf 65}, 066004 (2002)
  [arXiv:hep-th/0105137].
}

\lref\LSmodel{
  M.~Luty and R.~Sundrum,
  ``Anomaly mediated supersymmetry breaking in four dimensions,  naturally,''
  Phys.\ Rev.\  D {\bf 67}, 045007 (2003)
  [arXiv:hep-th/0111231].
}

\lref\ypq{
  J.~P.~Gauntlett, D.~Martelli, J.~Sparks and D.~Waldram,
  ``Sasaki-Einstein metrics on S(2) x S(3),''
  Adv.\ Theor.\ Math.\ Phys.\  {\bf 8}, 711 (2004)
  [arXiv:hep-th/0403002].
}

\lref\lpqr{
  M.~Cvetic, H.~Lu, D.~N.~Page and C.~N.~Pope,
  ``New Einstein-Sasaki spaces in five and higher dimensions,''
  Phys.\ Rev.\ Lett.\  {\bf 95}, 071101 (2005)
  [arXiv:hep-th/0504225].
}

\lref\IS{
  K.~Intriligator and N.~Seiberg,
  ``The runaway quiver,''
  JHEP {\bf 0602}, 031 (2006)
  [arXiv:hep-th/0512347].
}

\lref\Cvetic{
  R.~Blumenhagen, M.~Cvetic, P.~Langacker and G.~Shiu,
  ``Toward realistic intersecting D-brane models,''
  Ann.\ Rev.\ Nucl.\ Part.\ Sci.\  {\bf 55}, 71 (2005)
  [arXiv:hep-th/0502005].
}

\lref\Marchesano{
  F.~Marchesano,
  ``Progress in D-brane model building,''
  arXiv:hep-th/0702094.
}

\lref\dpcasc{
  S.~Franco, A.~Hanany and A.~M.~Uranga,
  ``Multi-flux warped throats and cascading gauge theories,''
  JHEP {\bf 0509}, 028 (2005)
  [arXiv:hep-th/0502113].
}

\lref\ypqcasc{
  C.~P.~Herzog, Q.~J.~Ejaz and I.~R.~Klebanov,
  ``Cascading RG flows from new Sasaki-Einstein manifolds,''
  JHEP {\bf 0502}, 009 (2005)
  [arXiv:hep-th/0412193].
}

\lref\bertolini{
  M.~Bertolini, F.~Bigazzi and A.~L.~Cotrone,
  ``Supersymmetry breaking at the end of a cascade of Seiberg dualities,''
  Phys.\ Rev.\  D {\bf 72}, 061902 (2005)
  [arXiv:hep-th/0505055].
}

\lref\berenstein{
  D.~Berenstein, C.~P.~Herzog, P.~Ouyang and S.~Pinansky,
  ``Supersymmetry breaking from a Calabi-Yau singularity,''
  JHEP {\bf 0509}, 084 (2005)
  [arXiv:hep-th/0505029].
}

\lref\Franco{
  S.~Franco, A.~Hanany, F.~Saad and A.~M.~Uranga,
  ``Fractional branes and dynamical supersymmetry breaking,''
  JHEP {\bf 0601}, 011 (2006)
  [arXiv:hep-th/0505040].
}

\lref\diaconescu{
  D.~E.~Diaconescu, R.~Donagi and B.~Florea,
  ``Metastable quivers in string compactifications,''
  arXiv:hep-th/0701104.
}

\lref\Argurio{
  R.~Argurio, M.~Bertolini, C.~Closset and S.~Cremonesi,
  ``On stable non-supersymmetric vacua at the bottom of cascading theories,''
  JHEP {\bf 0609}, 030 (2006)
  [arXiv:hep-th/0606175].
}

\lref\Uranga{
  A.~M.~Uranga,
  ``Brane configurations for branes at conifolds,''
  JHEP {\bf 9901}, 022 (1999)
  [arXiv:hep-th/9811004].
}

\lref\WS{
  C.~P.~Burgess, P.~G.~Camara, S.~P.~de Alwis, S.~B.~Giddings, A.~Maharana, F.~Quevedo and K.~Suruliz,
  ``Warped supersymmetry breaking,''
  arXiv:hep-th/0610255.
}

\lref\KS{
  I.~R.~Klebanov and M.~J.~Strassler,
  ``Supergravity and a confining gauge theory: Duality cascades and
  chiSB-resolution of naked singularities,''
  JHEP {\bf 0008}, 052 (2000)
  [arXiv:hep-th/0007191].
}

\lref\KW{
  I.~R.~Klebanov and E.~Witten,
  ``Superconformal field theory on threebranes at a Calabi-Yau  singularity,''
  Nucl.\ Phys.\  B {\bf 536}, 199 (1998)
  [arXiv:hep-th/9807080].
}

\lref\ibe{M. Ibe, K. Izawa, Y. Nakayama, Y. Shinbara and T. Yanagida,
``Conformally sequestered SUSY breaking in vector-like gauge theories,''
Phys.\ Rev.\ D {\bf 73}, 015004 (2006) [arXiv:hep-ph/0506023]\semi
M. Ibe, K. Izawa, Y. Nakayama, Y. Shinbara and T. Yanagida,
``More on conformally sequestered SUSY breaking,''
Phys.\ Rev.\ D {\bf 73}, 035012 (2006) [arXiv:hep-ph/0509229].}

\lref\gaugino{
  D.~E.~Kaplan, G.~D.~Kribs and M.~Schmaltz,
  ``Supersymmetry breaking through transparent extra dimensions,''
  Phys.\ Rev.\  D {\bf 62}, 035010 (2000)
  [arXiv:hep-ph/9911293]\semi
  Z.~Chacko, M.~A.~Luty, A.~E.~Nelson and E.~Ponton,
  ``Gaugino mediated supersymmetry breaking,''
  JHEP {\bf 0001}, 003 (2000)
  [arXiv:hep-ph/9911323].
}

\lref\KKLMMT{
  S.~Kachru, R.~Kallosh, A.~Linde, J.~M.~Maldacena, L.~McAllister and S.~P.~Trivedi,
  ``Towards inflation in string theory,''
  JCAP {\bf 0310}, 013 (2003)
  [arXiv:hep-th/0308055].
}

\lref\Baumann{
  D.~Baumann, A.~Dymarsky, I.~R.~Klebanov, J.~Maldacena, L.~McAllister and A.~Murugan,
  ``On D3-brane potentials in compactifications with fluxes and wrapped
  D-branes,''
  JHEP {\bf 0611}, 031 (2006)
  [arXiv:hep-th/0607050].
}

\lref\Ferrprobe{
  S.~Ferrara, M.~A.~Lledo and A.~Zaffaroni,
  ``Born-Infeld corrections to D3 brane action in AdS(5) x S(5) and N = 4,  d =
  4 primary superfields,''
  Phys.\ Rev.\  D {\bf 58}, 105029 (1998)
  [arXiv:hep-th/9805082].
}

\lref\Kabsorb{
  S.~S.~Gubser, A.~Hashimoto, I.~R.~Klebanov and M.~Krasnitz,
  ``Scalar absorption and the breaking of the world volume conformal
  invariance,''
  Nucl.\ Phys.\  B {\bf 526}, 393 (1998)
  [arXiv:hep-th/9803023].
}

\lref\spectrum{
  A.~Ceresole, G.~Dall'Agata, R.~D'Auria and S.~Ferrara,
  ``Spectrum of type IIB supergravity on AdS(5) x T(11): Predictions on N  = 1
  SCFT's,''
  Phys.\ Rev.\  D {\bf 61}, 066001 (2000)
  [arXiv:hep-th/9905226]\semi
  A.~Ceresole, G.~Dall'Agata and R.~D'Auria,
  ``KK spectroscopy of type IIB supergravity on AdS(5) x T(11),''
  JHEP {\bf 9911}, 009 (1999)
  [arXiv:hep-th/9907216].
}

\lref\ofer{
  O.~Aharony, Y.~E.~Antebi and M.~Berkooz,
  ``Open string moduli in KKLT compactifications,''
  Phys.\ Rev.\  D {\bf 72}, 106009 (2005)
  [arXiv:hep-th/0508080].
}

\lref\DD{
  F.~Denef and M.~R.~Douglas,
  ``Distributions of flux vacua,''
  JHEP {\bf 0405}, 072 (2004)
  [arXiv:hep-th/0404116].
}

\lref\GKT{
  A.~Giryavets, S.~Kachru and P.~K.~Tripathy,
  ``On the taxonomy of flux vacua,''
  JHEP {\bf 0408}, 002 (2004)
  [arXiv:hep-th/0404243].
}

\lref\orbCFT{
  S.~Kachru and E.~Silverstein,
  ``4d conformal theories and strings on orbifolds,''
  Phys.\ Rev.\ Lett.\  {\bf 80}, 4855 (1998)
  [arXiv:hep-th/9802183].
}

\lref\LNV{
  A.~E.~Lawrence, N.~Nekrasov and C.~Vafa,
  ``On conformal field theories in four dimensions,''
  Nucl.\ Phys.\  B {\bf 533}, 199 (1998)
  [arXiv:hep-th/9803015].
}

\lref\ABFK{R. Argurio, M. Bertolini, S. Franco and S. Kachru, ``Gauge/gravity duality
and metastable dynamical supersymmetry breaking,'' arXiv:hep-th/0610212; and to appear.}

\noblackbox \Title{\vbox{\baselineskip12pt\hbox{PUPT-2226} \hbox{{\tt hep-th/0703105}}
\hbox{} }}
 {\vbox{ {\centerline{Sequestering in String Theory} }}}

\centerline{Shamit Kachru,$^{a}$ Liam McAllister,$^{b}$ and Raman
Sundrum$^{c}$}
\bigskip

{\it \centerline{$^{a}$SLAC and Department of Physics, Stanford
University, Stanford CA 94305}}

{\it \centerline{$^{b}$Department of Physics, Princeton
University, Princeton NJ 08540}}

{\it \centerline{$^{c}$Department of Physics and Astronomy, Johns
Hopkins University, Baltimore MD 21218 }}

\smallskip

\bigskip \noindent
We study sequestering, a prerequisite for flavor-blind supersymmetry
breaking in several high-scale mediation mechanisms,
in compactifications of type IIB string theory. We find that although sequestering is typically absent in
unwarped backgrounds, strongly warped compactifications do
readily sequester.  The AdS/CFT dual description in terms of conformal sequestering plays an important role in our analysis, and we 
establish how sequestering works both on the gravity side and on the gauge theory side.
We pay special attention to subtle compactification effects that can disrupt sequestering.  Our result is a step toward realizing an appealing pattern of soft terms in a KKLT compactification.

\Date{March 2007}
\listtoc \writetoc

\newsec{Introduction}

Several of the most promising ideas about high-scale transmission of
supersymmetry (SUSY) breaking to the supersymmetric standard model
(SM) involve the physics of extra dimensions. These ideas, which
include anomaly mediation \refs{\RS,\other}, gaugino mediation
\gaugino, and mirage mediation \mirage, each require the strong suppression of
standard gravity-mediation effects, which would otherwise naively dominate.
For this to happen robustly, it is important that
the SUSY-breaking hidden sector be ``sequestered'' away from the visible
SM fields in the extra-dimensional space.

Concretely, the troubling gravity-mediation term is really a direct
cross-coupling in
the effective Lagrangian written in curved superspace,
between the hidden-sector fields $X$ bearing the dominant
F-term, and the visible-sector fields $Q^i$, of the form
\eqn\dangerous{\int d^4\theta ~X^\dagger X ~{c_{ij}\over
M_P^2} (Q^i)^\dagger Q^j~.} For order-one and flavor-violating $c_{ij}$,
phenomenologically
dangerous flavor-violating squark and slepton mass terms would result.
This is a version of the ``supersymmetric flavor problem''.
We say that $X$ and $Q$ have been
sequestered from one another if $c_{ij} \ll 1$.
It is important to note \RS\ that the cross-couplings in curved superspace
should be identified with  cross-couplings
 within \dangerous\ in \eqn\deff{f = - 3M_P^2
~{\rm exp}\left(-{K \over 3M_P^2} \right)} rather than within the
effective four-dimensional supergravity K\"ahler potential, $K$, itself.

As originally envisioned, sequestering could be justified by
locality in extra dimensions of space \RS. That is, if SUSY
were broken on a brane separated from the SM brane by a distance
$d$ in the extra dimensions, cross-couplings in the four-dimensional effective theory
should only arise from integrating out the exchange of massive bulk modes.
If these modes all had
sufficiently large masses $m \gg 1/d$, these bulk exchanges would be
 exponentially screened by Yukawa suppression, leading to
$c_{ij} \sim e^{- m \cdot d} \ll 1$.
In this circumstance, other effects, {\it{e.g.}} the
anomaly mediated contribution $\sim {1\over 16\pi^2}g^2 {|F|^2
\over M_P^2}$, could provide dominant, and flavor-blind, soft
masses.\foot{For the particular case of anomaly mediation,
additional mechanisms are required to avoid tachyonic sleptons,
with concrete suggestions appearing in {\it{e.g.}}
\refs{\RS,\Pomarol}.} An important subtlety is that at least some bulk modes
should get masses $m \lesssim 1/d$ from compactification itself or from moduli stabilization, 
and therefore
their exchanges require careful analysis. A thorough study was made
within a minimal five-dimensional  effective field theory with a single extra dimension,
confirming that sequestering was indeed robust in that context \LS.

Nevertheless, it has been argued that in string theory, with its much
richer structure and several extra dimensions,
 it is very difficult to
realize  sequestering
\refs{\Anisimov,\KMS}.
  In particular, these works made it clear
that spatial separation alone does {\it{not}} suffice to ensure
sequestering in string constructions.  However,
these studies were not exhaustive and whether
sequestering could be realized in a complete theory of quantum gravity
remained an open and important question.

A different approach to sequestering was initiated in \Ls.
It was shown there, again within minimal five-dimensional effective field theory, that
sequestering was consistent with strongly warped compactifications and
their attendant geometric hierarchies. This suggested, via the AdS/CFT
correspondence \refs{\Maldacena,\OtherAdSCFT}, that there should exist a dual four-dimensional mechanism for sequestering.
In ``conformal sequestering'' \LSconf, instead of a weakly coupled
hidden sector, we imagine a hidden sector that is strongly coupled
over some range of scales, and is close to a conformal fixed point
over the range from $\Lambda_{UV}$ to $\Lambda_{IR}$.
(An alternate
approach of conformal sequestering,
where it is the visible sector that is strongly coupled
and conformal over a range of scales, was proposed earlier in \Nelson.)
  At the
scale $\Lambda_{IR}$, conformal invariance breaks, and soon after
this, spontaneous supersymmetry breaking occurs. Some operator
${\cal O}$ in the hidden sector CFT provides the leading coupling
\eqn\leading{ \int d^4\theta ~{\cal O}~c_{ij}(Q^i)^\dagger Q^j} of
SUSY breaking to the visible squarks and sleptons.
But now if ${\cal O}$ has scaling
dimension $2+\Delta$, one finds an additional suppression of the
naive coupling \dangerous\ by a factor of $\left(\Lambda_{IR}\over
\Lambda_{UV}\right)^{\Delta}$. For a typical intermediate-scale
scenario, this suppresses the dangerous cross-couplings
sufficiently for anomaly mediation (or one of the other
extra-dimensional mechanisms) to provide safely flavor-universal visible
masses. Conformal sequestering has been further developed in \refs{\LSmodel,\ibe,\SS}.

In this paper we demonstrate that the mechanism
 of ``warped sequestering'' can be realized in string theory, in the
form leading to anomaly-mediated supersymmetry breaking in the visible sector.
In essence, we will describe how warped
compactifications of string theory, involving warped throats
similar to the canonical Klebanov-Strassler example \KS,
compactified as in \GKP, can naturally yield sequestered SUSY
breaking. We will make considerable use of the AdS/CFT dual grammar
of conformal sequestering, although the explicit construction will be
on the gravity side of the duality. We also show that
sequestering is possible in unwarped compactifications, but typically
absent.

Our goal here is to establish that sequestering can be achieved fairly
robustly in a theory of quantum gravity. We will not discuss the
construction of fully realistic models. This would require both
inclusion of a detailed visible sector, and incorporation of one of the
various model-building mechanisms to {\it{e.g.}} fix the tachyonic
sleptons of anomaly mediation.

We will restrict ourselves to type IIB
string theory constructions, but
our final conclusions will clearly be more general. More precisely,
since we will always consider non-vanishing brane separations to be
 much
larger than the string length, we will work in terms of type IIB effective
supergravity coupled to branes.

\subsec{A milder criterion for sequestering}

While sequestering originally referred to suppression of {\it all}
 terms of the form \dangerous, we will make use of a milder criterion
following from a simple observation of \SS.  This states that
 we can tolerate an unsuppressed
bilinear hidden superfield appearing in \dangerous\ if it
 has as its vector component a conserved
Noether current for a hidden sector IR (non-R)
symmetry,  because then the hidden bilinear is guaranteed to have
 vanishing SUSY-breaking VEV.
The symmetry is only required to be good symmetry of the hidden
dynamics close to the SUSY breaking scale, {\it even if it is
ultimately spontaneously broken by hidden VEVs}.

We discuss here a simple
 example of the power of this observation, and will have occasion to
return to this example in the course of the paper. It
 is provided by
considering a distribution of D3-branes in a
simple toroidal orientifold $T^6/Z_2$ compactification of the six extra dimensions,
preserving ${\cal N}=4$ supersymmetry. Splitting the branes into two well-separated
stacks, with chiral matter (in ${\cal N}=1$ language)
labeled $Q$ and $X$, provides a toy model of visible and hidden sectors.
While this $X$ hidden sector does not break SUSY, it does allow us to
study sequestering itself. This example was previously invoked as a
case in which string theory does not sequester \Anisimov, as follows.
The K\"ahler potential is readily determined
by the high degree of supersymmetry and turns out to be
\eqn\torus{f = -3 \Bigl[(S+S^\dagger) \det{ \left(T_{IJ} + T^{\dagger}_{JI} -
{\rm{tr}}\, Q_I Q^\dagger_J - {\rm{tr}}\, X_I X^\dagger_J\right)} \Bigr]^{1/3} \, ,} where $I, J = 1,2,3$
label the ${\cal N}=1$ chiral superfields in an ${\cal N}=4$ multiplet and the
traces are over the different branes in each stack. We assume
that the moduli $S, T_{IJ}$ are supersymmetrically stabilized by
some contributions to their superpotential (while this would not happen
in the ${\cal N}=4$ theory, it can happen in close relatives that inherit
the relevant pieces of $f$). At low energies
they can therefore be set to their SUSY-preserving VEVs. It is clear
that expanding $f$ in powers of brane fields will result in cross-couplings
of the form \dangerous.

But $f$ does satisfy the milder version of
sequestering, as we now check.
 Expanding \torus\ in powers of
$Q, X$ (setting the moduli to their VEVs), we get 
\eqn\expandtorus{
f = const. - {\rm{tr}}\, \tilde{Q}^\dagger_I \tilde{Q}^I - {\rm{tr}}\, \tilde{X}^\dagger_I
\tilde{X}^I
+ {\rm{tr}}\, \tilde{Q}^\dagger_J \tilde{Q}^I  \left[\delta_{I}^{~J}
{\rm{tr}}\, \tilde{X}^\dagger_K \tilde{X}^K - 3\, {\rm{tr}}\,
 \tilde{X}^\dagger_I \tilde{X}^J\right]
+ \ldots,}
where the canonical fields are given by
\eqn\canontorus{\tilde{Q}_I = const. (T + T^\dagger)^{-1/2}_{IJ} Q_J~,}
\eqn\canontwo{
\tilde{X}_I = const. (T + T^\dagger)^{-1/2}_{IJ} X_J.}
The hidden bilinear, traceless in the $I, J$ indices,
 manifestly corresponds to currents of the $SU(3)$ subgroup of
the ${\cal N}=4$ R-symmetry. From the ${\cal N}=1$ viewpoint, $SU(3)$ is a non-R flavor
symmetry acting on chiral multiplets. While the toroidal compactification
breaks this $SU(3)$ in the UV,
it is an accidental symmetry of the hidden sector in
the IR and therefore this example satisfies the milder criterion of \SS.
(Later we will obtain a more general understanding of why this happened.)

\subsec{Outline}

The remainder of the paper is organized as follows.
In \S2, we begin by studying unwarped backgrounds with D-branes,
ultimately specializing to D3-branes. We show how unwarped compactification
in general spoils sequestering, although we point out exceptional cases where
sequestering holds to a good approximation.
In \S3, we study branes in highly warped backgrounds, in particular
specializing to the case of the Klebanov-Strassler conifold throat. We
discuss the AdS/CFT dual description of conformal sequestering and
show that, even upon compactification, sequestering must hold. In \S4
we illustrate
some aspects of this sequestering directly from computations on the
gravity side of the duality. We illustrate the conjunction of both hidden
sector SUSY breaking and warped sequestering by the example of an anti-D3-brane at the end of the throat \KPV. In \S5 we discuss general classes and
properties of warped throats which appear to be promising for warped
sequestering. We conclude in \S6.

\newsec{Sequestering in unwarped backgrounds}

In this section we discuss the possibility of sequestering in compactifications of type IIB string theory with little or no warping.  Although we will eventually be led to pursue sequestering in warped compactifications, the unwarped case provides an instructive introduction to the general problem. In \S2.1 we study D-branes coupled to supergravity in the simplest context, ten non-compact dimensions.  In particular, we explain why the fields on spatially-separated D3-branes might be expected to enjoy sequestering. Then, in \S2.2, we expose the fallacy in this logic: compactification effects do generically spoil sequestering of D3-branes in unwarped compactifications, albeit with some notable exceptions.  We then describe the effects on sequestering from K\"ahler moduli corresponding to blowup modes (\S2.3), from complex structure moduli (\S2.4), and from quantum corrections to the K\"ahler potential (\S2.5).

\subsec{Restrictions on D3-brane couplings}

We begin by recalling, from \Anisimov, a reason for fearing that two stacks of D-branes, say stack Q and stack X, will
have dangerous cross-terms such as \dangerous\ in their four-dimensional effective Lagrangian.  In string theory, D-branes source various ten-dimensional fields, such as the metric and dilaton, and thereby affect their surroundings.  By affecting the background in this way, each stack can perceive the other.  Specifically, denote the metric on the compactification manifold
by $g_{mn}(y)$, and the axio-dilaton as $\tau(y)$. Then backreaction of stack $Q$ on $g_{mn}$ and $\tau$,
which will implicitly make $g$ and $\tau$ into functions of the scalar
modes of $Q$, will translate directly into couplings between
$Q$ and $X$ when one computes the probe action for the $X$
branes in the same background.

As a concrete example, we can consider the toy model on p.40 of the second reference of \Anisimov.
Imagine that the X-stack is the hidden sector, and is
composed of a Dp$^\prime$ brane.  Let the Q-stack, where the visible sector
resides, be the probe brane.  The Dp$^\prime$ brane background
metric and dilaton are given by \eqn\backA{ds^2 = h(r)^{-1/2}
dx_{||}^2 + h(r)^{1/2} dx_{\perp}^2,~~ e^{-2\phi} =
h(r)^{(p'-3)/2}~} where the harmonic function $h(r)$ is given by \eqn\hff{h(r) = 1 +
g_s \left( {\sqrt{\alpha^\prime} \over r}\right)^{7-p'}~.}

The action of brane Q in this background is given by expanding the
DBI action \eqn\DBI{S_Q = -T_{Dp} \int d^{p+1}x~ e^{-\phi}
\sqrt{{\rm{det}}(\gamma_{\mu\nu} + \cdots)}~} where $\gamma_{\mu\nu}$ is the
induced world-volume metric, and the omitted terms involve worldvolume fluxes that are unimportant at present.
Expanding in the background \backA\ yields 
\eqn\DBIexp{ S_Q \propto \int d^{p+1}x~ \sqrt{{\rm{det}}(\tilde{\gamma}_{\mu\nu})}~h(r)^{(p'-p)/4-1}\left( 1 + {1\over 2} h(r) \partial_{\mu}\phi^i
\partial^{\mu}\phi_i + \ldots \right)~,}
where $\tilde{\gamma}_{\mu\nu}$ is the worldvolume metric induced from the
{\it{unwarped}} background metric.

Interestingly, when the solution takes the form \backA\ for
$p=p'$, the leading effect of the backreaction of the X-branes on
the Q-branes does not alter the kinetic terms.\foot{The appearance of
a potential in \DBI\ in this case is spurious: it is canceled by
the Chern-Simons terms in the action.}
This is the question of interest for sequestering, since the
dangerous term \dangerous\ would imply $|X|^2 |\partial Q|^2$
scalar cross-couplings (among other terms).

This example is at best a toy model for the full conformally
Calabi-Yau solutions with nontrivial fluxes of the NSNS three-form
$H_3$ and the RR three-form $F_3$ and five-form $F_5$. The general supergravity solution in the presence of these
fluxes is discussed in some detail in \GKP\ (building on the earlier
works \refs{\Becker,\GVW,\DRS}).  The main points, at
leading order in the $g_s$ and $\alpha^\prime$ expansions, are the
following:

\noindent $\bullet$ The metric takes the form \eqn\metis{ds^2 =
e^{2A(y)} \eta_{\mu\nu}dx^{\mu}dx^{\nu} + e^{-2A(y)} \tilde
g_{mn}(y) dy^m dy^n} where the warp factor $A(y)$ depends on
the D3-brane positions and the D7-brane positions, as well as on the
background flux.

\noindent $\bullet$ The dilaton takes the form $\tau=\tau(y)$, and
is determined self-consistently in terms of $\tilde g_{mn}$ and
the positions of any D7-branes.  There is {\it no} explicit
dependence on the D3-brane positions.

\noindent $\bullet$ The Einstein equations determining the
internal metric $\tilde g_{mn}$ depend on the dilaton gradients
and the localized D7-brane actions, but again have {\it no}
explicit dependence on the D3-brane positions.

It follows from these facts that the cancellation that prevents
cross-couplings of the form \dangerous\
 between D3-brane fields from different stacks in the
simple ansatz \backA\ will continue to hold for the probe analysis
of kinetic terms of a given D3-brane stack in the full
supergravity solutions of \GKP. While the D3-branes see any
background D7-branes via their kinetic terms, they do not see each
other at this order. In fact, the D3-branes also enjoy a vanishing
potential, even though generic fluxes do break supersymmetry.

We note that this reasoning does not work for inter-D7-brane couplings
in the same class of type IIB solutions. The fortuitous
cancellation of metric and dilaton factors that occurs for
D3-branes probing other D3-branes in the background \backA, does
not occur for general Dp-brane pairs.  It is a happy fact of
life for the D3-branes that their positions only enter the
supergravity data via factors of the warp factor $e^A$, and these
factors cancel in the leading terms of the probe action for a
D3-brane.

\subsec{Non-sequestered couplings via compactification moduli}

The fact that separated stacks of D3-branes are sequestered in non-compact ten-dimensional
spacetime might lead us to expect the same feature after compactification,
at least to leading order  in $\alpha'$ and
$g_s$. However,
we now study the four-dimensional supergravity theory after compactification and show that it
typically  leads to complete breakdown of sequestering in backgrounds of the form of \S2.1.
While we limit ourselves in this section to Calabi-Yau
orientifolds with D3-branes, incorporating D7-branes would
not be difficult. Some of the results here were described (or
are implicit) in
\refs{\GKP,\Anisimov,\louisone,\DeWolfe,\KKLMMT,\louistwo,\GM,\Baumann}.

The loophole in our previous discussion is that compactification results in moduli  whose
parametrization in terms of ${\cal N}=1$
chiral superfields depends on the locations of any D3-branes. Couplings of these moduli
to the D3-branes can thereby mediate non-sequestered cross-couplings between
distant branes. We begin by explaining this in the case of a single breathing mode for the
entire compact space.

We now use the ansatz \eqn\lineel{ ds^2 =
h^{-1/2}e^{-6u}g_{\mu\nu}dx^{\mu}dx^{\nu} +
h^{1/2}e^{2u}\tilde{g}_{ij}dy^i dy^j \, .} Here $h(y) = e^{-4A(y)}$ is
the usual warp factor, $e^{6u}$ is the breathing mode of the
compactification manifold $M$, and ${\tilde{g}}$ is the unwarped
metric, with fixed fiducial volume $\tilde{V}_6$.  The powers of
$e^u$ are chosen so that $g_{\mu\nu}$ is the spacetime metric in
four-dimensional Einstein frame.

We will now show (following \Baumann) that the breathing mode depends on the D3-brane
positions.  We will then demonstrate that this leads to contact terms between separated
D3-branes.  The perturbation to the warp factor, $\delta h$, sourced by a single D3-brane
at $x$ obeys \eqn\laplace{ -\nabla_{y}^2 \delta h(x;y) =
2\kappa_{10}^2 T_3
\biggl({{{\delta^{(6)}(x-y)}\over{\sqrt{g}}}-\rho(y)} \biggr)\, ,}
where $\rho(y)$ is the background density of D3-brane charge,
arising from fluxes and possibly from other branes.  Here $y$ is
the coordinate on the internal space, and $x$ is the position of
the D3-brane on this space.  The distinction is important: the
four-dimensional action arising from dimensional reduction does
not depend on $y$, but $x$ is a scalar field in this action --- $x$
is a coordinate on the D3-brane moduli space.  (This moduli space
has the same metric as the physical compact space, at least at
leading order.)  Finally, the D3-brane tension is \eqn\dthreetension{ T_3 = {1\over{(2\pi)^3 g_s {\alpha^{\prime}}^2}}\, ,}
and the ten-dimensional gravitational coupling is \eqn\kappaten{ \kappa_{10}^2 = {1\over{2}}(2\pi)^7 g_s^2 {\alpha^{\prime}}^4 \, .}

We solve \laplace\ by first solving \eqn\easylaplace{
-\nabla_{y'}^2 \Phi(y;y') = - \nabla_y^2 \Phi(y;y') =
{{\delta^{(6)}(y-y')}\over{\sqrt{g}}}-{1\over{{V_6}}}} where $V_6
\equiv \int d^6 y \sqrt{g} = e^{6u}\tilde{V}_6$. The solution to
\laplace\ is then \eqn\solvedlaplace{ \delta h(x;y) = 2\kappa_{10}^2
T_3 \Bigl[ \Phi(x;y) - \int d^6 y' \sqrt{g}\, \Phi(y;y') \rho(y')
\Bigr] \, .}  An important consequence \Baumann\ is that \eqn\modulicharge{ -
\nabla_x^2 \delta h(x;y) = 2\kappa_{10}^2 T_3
\left[{\delta^{(6)}(x-y) \over \sqrt{g(x)}} - {1\over{{V_6}}}
\right] \, ,} for any background charge distribution $\rho(y)$.  Away from the D3-brane we may drop the $\delta^{(6)}$ term in
\modulicharge, which leads to
\eqn\simple{\tilde{g}^{i\bar{j}}\partial_{i}\partial_{\bar{j}}\delta
h = \left({{\kappa_{10}^2 T_3}\over{\tilde{V}_6}}\right)
e^{-4u}\equiv {T_3\over{M_{P}^2}} e^{-4u} \, ,} where the power of $u$
comes from converting $V_6$ to $\tilde{V}_6$ and $g^{ij}$ to
$\tilde{g}^{ij}$, and we have assumed that $M$ is K\"ahler to
simplify the Laplacian to the form shown.

The right hand side of \simple\ is a constant independent of $y$,
but $\tilde{g}^{i\bar{j}}$ depends on $y$.  This is consistent
only if $\delta h$ obeys \eqn\inverse{
\partial_{i}\partial_{\bar{j}}\delta h =
{T_3\over{3 M_{P}^2}} e^{-4u} \tilde{g}_{i\bar{j}} \, .} We
conclude that (to leading order in the gravitational coupling \Baumann) \eqn\delis{ \delta h = {T_3\over{3 M_{P}^2}} e^{-4u} k(x,\bar{x}) + \ldots \, ,} where $k$ is the
K\"ahler potential for the metric $\tilde{g}$ on $M$, and the dots
denote terms annihilated by the Laplacian, which are not
constrained by the above argument.

Next, we use $h + \delta h$
to identify the appropriate holomorphic coordinate $T$ on
the K\"ahler moduli space.  As explained in detail in
\refs{\GM,\Baumann}, the appropriate definition that is consistent
with \delis\ is 
\eqn\nowtis{ T +{\overline{T}} \equiv \int_{\Sigma} d^4 y \sqrt{G} = \int_{\Sigma} d^4 y \sqrt{\tilde{g}}\,h e^{4u} \, ,}
where $G_{ij} \equiv h^{1/2}\tilde{g}_{ij}e^{2u}$ is the full metric on the compact space.  Here $\Sigma$ is an appropriate four-cycle, and because we have one
K\"ahler modulus, $\Sigma$ is effectively unique.  We may
normalize the fiducial metric $\tilde{g}$ so that \eqn\fiducial{
\int_{\Sigma} d^4 y \sqrt{\tilde{g}} = 1 \, .}  Next, we choose for
simplicity a case without background warping, so that $h=1+\delta
h$ gets its only nontrivial contribution from the D3-branes in
question.  Combining these ingredients with \delis, we find 
\eqn\defoft{ T +{\overline{T}} = e^{4u} +\gamma k(x,\bar{x}) \, ,} where
\eqn\gammais{ \gamma \equiv {T_3 \over{3 M_{P}^2}},}
which makes manifest the D3-brane-dependence of
the holomorphic parametrization of the breathing mode.
(As discussed in \KKLMMT, invariance of this expression under K\"ahler
transformations of $k$ requires that $T$ also transforms.)
It follows (using the standard relation between the K\"ahler potential
for volume moduli and the compactification volume)
that the K\"ahler potential is \eqn\nowkis{ K
= - 3~ {\rm{log}}\Bigl( T + \overline{T} - \gamma k(x,\bar{x}) \Bigr) \, .}
By repeating the above argument for an additional D3-brane located
at $x^{\prime}$, we find \eqn\kwithtwobrane{ K = - 3~
{\rm{log}}\Bigl( T + \overline{T} - \gamma k(x,\bar{x})-\gamma
k(x^{\prime},\bar{x}^{\prime}) \Bigr).}

Notice that this
result is sequestered: the coefficient of $3$ implies
that the different brane stacks and $T$ appear additively in $f$.
This feature will, however, turn out to be a special case, as we now demonstrate
by extending the above considerations to a configuration with
multiple K\"ahler moduli and two stacks of D3-branes.  The condition
\modulicharge\ is unchanged, as it depends only on the overall
breathing mode, not the relative sizes of cycles.  It follows that
\delis\ is also unchanged.  Next, if $\{\Sigma_{\alpha}\}$ is a basis of four-cycles,
we can identify the proper holomorphic coordinates $T_{\alpha}$, \eqn\talpha{ T_{\alpha}
+\overline{T}_{\alpha} \equiv \int_{\Sigma_{\alpha}}
\sqrt{\tilde{g}}\,e^{4u}\,\Bigl(1+\delta h(x,x')\Bigr) \, .}
The K\"ahler-invariant coordinates are then \eqn\ualpha{ U_{\alpha} =
T_{\alpha} + \overline{T}_{\alpha} - \gamma
k_{\alpha}(x,\bar{x})-\gamma
k_{\alpha}(x^{\prime},\bar{x}^{\prime})} 
where the transformation properties of $k_{\alpha} \sim k \int_{\Sigma_{\alpha}} \sqrt{\tilde{g}}$ are dictated by \Ganor.
The K\"ahler potential is then \eqn\totalk{ {{K}} = - 2 ~{\rm{log}}(V_6)
= - 2 ~{\rm{log}}\Bigl(
d^{\alpha\beta\gamma}U_{\alpha}^{1/2}U_{\beta}^{1/2}U_{\gamma}^{1/2}
\Bigr)} where $d^{\alpha\beta\gamma}$ is determined 
by the intersection form of the four-cycles $\{\Sigma_{\alpha}\}$.

From \ualpha, \kwithtwobrane\ it is easy to see that
non-sequestered terms between the D3-branes are indeed generated in
compactifications with multiple K\"ahler moduli, but not in the
single-modulus case.

As we discussed in \S1\ and will revisit in \S4.3,
the ${\cal N}=4$ toroidal orientifold has multiple K\"ahler moduli
but provides an interesting exception to the rule: this
configuration exhibits sequestering in the subtler sense of \SS.
However, general multi-moduli compactifications
are not expected to be safe in this sense.

Our argument for \ualpha\ ({\it{cf.}} \Baumann) was slightly
indirect, involving the background charge in \laplace\ and the
associated transformation of the holomorphic coordinates
$T_{\alpha}$ under K\"ahler transformations.  However, an earlier
and far more explicit derivation of this same expression appears
in \louisone. Nevertheless, our method is instructive: we have
seen that the mechanism by which D3-branes couple to each other in $f$ is
by mixing with the holomorphic K\"ahler moduli $T_{\alpha}$, and,
crucially, this mixing is a result of the {\it{compactness}} of
the space, as it follows from the Gauss's law constraint \modulicharge.

\subsec{Sequestering in the case of ``localized''  K\"ahler moduli}

The explicit results of \louisone\ suggest that sequestering is
spoiled in typical models with multiple K\"ahler moduli.  Here we point
out the existence of a class of multi-moduli models in which approximate sequestering can arise.

Consider the following  example with two K\"ahler moduli.
The
Calabi-Yau hypersurface in ${\rm{W}}\IP^{4}_{1,1,1,6,9}$  has (as
derived in {\it{e.g.}} \Quevedo)
\eqn\calktwo{{K} \propto -2 ~{\rm log} \left(U_1^{3/2} -
U_2^{3/2}\right)~.} Very roughly, one should think of $U_1$ as the
overall volume of the bulk Calabi-Yau manifold, and of $U_2$ as
controlling the volume of a blow-up mode for a $\IC^3/\Z_3$
singularity localized in the bulk. Resolving the singularity
introduces a finite-volume $\IP^2$ in the geometry, with a smooth
neighborhood locally modeled on the total space of the bundle
${\cal O}(-3) \to \IP^2$.
It is implausible, at least for small values of the blow-up mode,
that introducing this little ``bump'' on the Calabi-Yau space
should destroy sequestering between separated D3-branes, which one
would have expected to hold in the absence of this blow-up.  How can we see this in
detail?

As explained in \louisone, in particular in their expression (3.13),
in the general multi-moduli case the D3-brane fields appearing in
\ualpha\ are multiplied by the harmonic form corresponding to the
K\"ahler modulus $U_{\alpha}$ {\it evaluated at the position on the
compact manifold} of the given D3-brane stack.  Therefore, if the compact manifold is
 a large space with point-like singularities
controlled by several additional K\"ahler moduli, for generic D3-brane positions, the corresponding $U_{\alpha}$ will
{\it not} have significant dependence on the D3-brane positions.  That is, the
wavefunction of the K\"ahler mode has a negligible overlap with the
D3-brane.

In the two-modulus case introduced above, one
would then find a K\"ahler potential of the form
\eqn\kexp{{K} = -2 ~{\rm log}
\left(U_1^{3/2} - d\right)~}
where $d$ is nearly {\it independent} of
the D3-brane modes. This is still not of the sequestered form, but one
can do a systematic expansion in powers of $d/{\rm Vol}$,
in which the leading term shows sequestering, and in
which the corrections are parametrically small.  This is in accord with the intuition that an arbitrarily small blow-up
controlled by an additional K\"ahler modulus should not completely destroy the sequestering that is present in the single modulus case.

In the context of warped compactifications
an argument along these lines has been given in
\ChoiRecent, where it was argued that the wavefunction of the bulk
K\"ahler moduli has negligible overlap with the small three-cycle
at the tip of a Klebanov-Strassler throat, thereby resulting in
sequestering.  We will find that this argument is
not quite justified, but sequestering does nevertheless hold in the more
relaxed sense of \SS, as discussed in the introduction.

\subsec{Complex structure moduli}

We have not yet discussed the role of complex structure moduli, but we now show that
they do not alter sequestering. For instance, consider
a compactification with a single K\"ahler modulus, which, as we showed in \S2.2, enjoys
sequestering. In detail, the K\"ahler potential (at leading order in $\alpha'$ and $g_s$)
with complex structure moduli $z^{i}$, a K\"ahler
modulus corresponding to the harmonic two-form $\omega_{m\bar n}$,
axio-dilaton $\tau$, and D3-brane moduli $\phi^m$ is given by
\louisone\ \eqn\kis{ {K_{tot}\over{M_{P}^2}} = K_{CS}(z) - {\rm log}\Bigl(-i(\tau-\bar
\tau)\Bigr) - 2~{\rm log}\left({1\over 6}{\cal
K}(T,z,\phi)\right)~.} Here \eqn\kcsis{K_{CS}(z) = -{\rm
log}\left(-i \int_M \Omega \wedge \bar \Omega\right),}
with $\Omega$ the unique (up to scale) holomorphic three-form on the
Calabi-Yau space, and
\eqn\calk{-2~ {\rm log} ~{\cal K} = -3 ~{\rm log}~ {2\over
3}\left( T + \overline{T} + {3i\over 2\pi} \omega_{m\bar n}{\rm{Tr}}(\phi^m
\bar\phi^{\bar n}) + {3\over 8\pi} \left(\omega_{m\bar n}\bar
z^{i}(\bar\chi_{i})^{\bar n}_{l}{\rm{Tr}}(\phi^m\phi^l) +
h.c.\right)\right)~,} where $\chi_{i}$ is related to a
complex structure deformation of the compact manifold in a way that is not
important for our discussion.

We concluded that sequestering holds because
when the D3-brane fields $\phi^m$ are split into two stacks, $Q$ and $X$,
the dependences on $Q$ and $X$ appear additively in
\eqn\fagg{f
= -3 M_P^2~ {\rm exp}(-K_{tot}/3M_P^2) \, ,} without
cross-couplings between fields living on distinct D3-branes.

The reader might, however, wonder about the role of the complex structure
moduli $z^{i}$, which evidently couple directly to the
D3-brane fields $\phi$ in \calk. In flux models, the $z^{i}$
can be stabilized supersymmetrically by the fluxes at some high
scale $\sim {\alpha^\prime}/\sqrt{V_6}$.  Does integrating out these
fields then lead to dangerous cross-couplings of the D3-brane
fields?

The answer is no.  After appropriately rescaling the $\phi$
fields to obtain canonical normalization (by Taylor expanding the
logarithm and rescaling by the appropriate power of the volume),
the couplings we are concerned about look like \eqn\worry{\int
d^4\theta ~\left( {\bar z} \phi \phi + h.c.~\right) .} When we
integrate out the $z$ fields, it is easy to see that we will
introduce contact terms between $\phi$ fields on different
D3-branes, but these terms will be of higher dimension than
\dangerous. We can safely ignore these higher-derivative terms.

\subsec{Quantum corrections to sequestering}

Let us continue to focus on the model of a single K\"ahler modulus
discussed in the previous subsection.
 The exact K\"ahler potential receives non-sequestered
 $\alpha^\prime$ and $g_s$
corrections to \calk.
The leading correction inducing communication between separated
D3-branes was discussed in \Berg, based on explicit string loop
calculations in certain toroidal orientifolds. The relevant
operator takes the form \eqn\ops{ \Delta {K} = {c \over
{T+\overline{T}}} \left( {q(\phi,\bar\phi) \over {\tau-\bar\tau}}
\right)~} where $q$ is some local function of the D3-brane fields.
It is evident that through the function \fagg, this correction
gives rise to a contact term between fields on different
D3-branes, arising from \eqn\deltaf{\Delta f \sim {c \,g_s \over
{T+\overline{T}}} k(\phi,\bar\phi)q(\phi,\bar \phi)~,} with $k$ the
leading-order D3-brane K\"ahler potential.  However, as expected
for a string loop correction, this term is parametrically small at
weak string coupling.  (The power of $T+\overline{T}$ does not
represent a suppression, as this is part of the four-dimensional Planck mass suppression
in \dangerous\ which is by itself insufficient for sequestering.)

The constant $c$ in \ops,\deltaf\ is important.  Loop corrections
in higher-dimensional supergravity and string theory are
suppressed as in naive dimensional analysis, and in fact the value
of $c$ inferred from orientifold calculations \Berg, \eqn\ctiny{c
= {1\over 128 \pi^6}\, ,} indicates that sequestering is not only
a parametric fact in some such models, but can also be numerically
effective, even for reasonably large values of $g_s$.

\newsec{Warped sequestering and AdS/CFT}

In the previous section we found that in compactifications with little warping,
sequestering is in general possible only in models with a single K\"ahler modulus.\foot{The approximate sequestering of \S2.3 is present in
the limit in which all the four-cycles associated to additional K\"ahler moduli are blown down, which may be thought of as a one-modulus limit.}
The absence of sequestering, despite the extra-dimensional separation of visible and hidden brane
stacks by some distance $d$, is ultimately accounted for by a coupling mediated by
bulk KK modes, which we have integrated out in arriving at the four-dimensional effective
supergravity. Since such non-sequestered contributions are not seen from
purely five-dimensional analyses in which the fifth dimension
separates the hidden and
visible sectors \refs{\LS,\Ls}, the KK modes in question must be due to the other
extra dimensions, with some characteristic radius $\sim R$.  When all compactification scales
are comparable, the KK mediation is unsuppressed, as we found above except in special circumstances.
However, if there is a geometric hierarchy $d \gg R$, KK mediation should be Yukawa-suppressed $\sim e^{-d/R}$, and we
would expect sequestering to arise.  Warped throats provide a class of suitably
asymmetric compactifications, and are readily realizable in string theory.  In this section, we demonstrate that configurations of D3-branes separated along a warped throat naturally enjoy sequestering.  The AdS/CFT correspondence provides a valuable tool in
establishing this result, and also in making the connection to conformal sequestering.

We begin by describing one particular warped throat background of interest,
then establish sequestering from the perspective of the gauge
theory dual to this warped throat. We return to illustrate the central
elements from the gravity viewpoint. This realization of sequestering is
powerful enough to accommodate complex and realistic
visible sectors engineered in Calabi-Yau bulk portions of warped compactifications, not just the simple
gauge theories on D3-branes that we will consider here.  We will not, however, pursue visible-sector model
building in this paper; for the state of the art, see {\it{e.g.}} \refs{\Cvetic,\Marchesano}.

\subsec{The background}

Our setting is a type IIB warped flux compactification with the
metric \metis.  Such warped Calabi-Yau compactifications arise naturally in type IIB string theory, in
the presence of three-form fluxes in the internal space (and a
five-form flux that is determined by the warp factor).

A particularly simple and explicit example is due to Klebanov and
Strassler \KS.  We start with a conifold, a singular Calabi-Yau
space of the form \eqn\conifold{\sum_{i=1}^4 z_i^2 = 0} in
$\IC^4$. Upon deformation, {\it{i.e.}} adding a small constant
$\varepsilon$ to the RHS of \conifold), one finds two natural
three-cycles in the geometry: a three-sphere $A$ that shrinks in
the singular limit $\varepsilon \to 0$, and its dual three-cycle
$B$.  When these cycles are threaded by three-form fluxes
\eqn\fthree{\int_A F_3 = M,~~\int_B H_3 = -K} with $M \gg 1$ and
$K \gg 1$, one finds a simple supergravity solution with many
interesting properties. The flux superpotential deforms \conifold\
to an equation with an exponentially small constant on the RHS.
Far from the associated smoothed tip, the solution takes the form of a
warped metric \metis\ on the Calabi-Yau cone over the Einstein
manifold $T^{1,1}$,  with \eqn\his{h(r) = e^{-4A(r)} = {27 \pi
\over 4r^4}{\alpha^\prime}^2 g_s M K + \ldots \, ,} where the dots
denote known logarithmic corrections that are unimportant for our
considerations.

Focusing on the AdS-like portion of the metric, this is just a
spacetime of the familiar form \eqn\ads{ds^2 = {r^2\over R^2}
(-dt^2 + dx_i^2) + {R^2\over r^2} dr^2~ ,} with $r$ the radial
direction of the throat, and \eqn\pars{R^4 = {27 \over 4}\pi g_s N
{\alpha^\prime}^2,~~N \equiv KM~.} The minimal redshift in this throat
arises at the smoothed tip (where the metric, although well-known
\KS, departs from the form we have shown here), and is \GKP\
\eqn\minis{{r_0\over R} = {\rm exp}\left(-{2\pi K
\over{3g_sM}}\right)~.} Further, we can glue  this throat
region into a compact Calabi-Yau around $r \approx r_{max}$ for
some large $r_{max}$, as in \GKP.

The field theory dual of the Klebanov-Strassler solution is an
$SU(N+M) \times SU(N)$ gauge theory with chiral multiplets
$A^{1,2}$ and $B^{1,2}$ in the representations $(N+M,
\overline{N})$ and $(\overline{N+M},N)$, respectively. The theory
has a superpotential \eqn\ksup{W = \epsilon_{ij}\epsilon_{kl} {\rm{Tr}}
A^i B^k A^j B^l } which is invariant under an $SU(2) \times SU(2)$
global symmetry, where one factor rotates the $A_i$ and the other
rotates the $B_j$. This symmetry is broken in the UV
by the compactification, but it is an accidental symmetry in the IR of the
CFT, or equivalently far down the throat.

If the fixed point were a free field theory, \ksup\ would be
irrelevant.  Instead, we should interpret this theory as follows.
For $M=0$, this theory is a CFT, dual to the
supergravity background created by N D3-branes at a conifold
singularity \KW.  The chiral fields $A, B$ have conformal
dimension 3/4 at the nontrivial fixed point.  The theory with
finite $M \ll N$ should be viewed in a $1/N$ expansion about this
interacting fixed point.  This results, as argued convincingly in
\KS, in a ``renormalization group cascade."  If $N=KM$, this
cascade occurs over a range of energy scales with
\eqn\range{{\Lambda_{IR}\over{\Lambda_{UV}}} = {\rm
exp}\left(-{2\pi K \over{3g_sM}}\right)~,} corresponding precisely
to \minis.

\subsec{Conformal sequestering}

Let us imagine that we can modify the background described in \S3.1 in such a
way that in the IR, at the scale $\Lambda_{IR}$, SUSY is broken.
Various ways of doing this in the dual gravity solution have been
proposed in {\it{e.g.}} \refs{\KPV,\KKLT}, and we will discuss
these in \S4. The main important fact will be that the leading
operators generating soft terms are highly irrelevant. Whether or not
the SUSY breaking spontaneously breaks
the $SU(2) \times SU(2)$ symmetry of the Klebanov-Strassler sector
(conifold throat), the SUSY breaking will be sequestered from a visible
sector localized elsewhere on the Calabi-Yau, as we now explain.

We follow the discussion in \SS.
Quite generally,  the leading correction to
sequestering will be of the form
\eqn\leadpert{\int d^4\theta
~\widehat {\cal O}~ Q^\dagger Q} where $\widehat {\cal O}$ is the hidden sector {\it
non-chiral operator with the smallest conformal dimension}.
In practice, a dimension $\Delta \gtrsim 3$  provides adequate sequestering.
If the hidden CFT has a non-R symmetry, there will be dimension-two non-chiral
operators in the same supermultiplet as the associated Noether currents
\refs{\LSconf,\DFG},
but as mentioned in the introduction, these operators will be harmless if
the relevant CFT deformations (that enable SUSY breaking) also preserve
this symmetry \SS.  (The
$U(1)_R$ itself is harmless, never mediating problematic
interactions; it is dual to the bulk graviphoton, which is
similarly harmless \LS).
The Klebanov-Strassler field theory is a deformation
of the Klebanov-Witten \KW\ CFT by a small number of field theory colors, and
does not alter the $SU(2) \times SU(2)$
global symmetry, so we conclude that there is no danger from the associated non-chiral operators.

The leading potentially dangerous non-chiral operators
are then determined by the properties of the
conifold throat in the gravity dual.
The spectrum of bulk KK modes in $T^{1,1}$
compactification of type IIB supergravity was determined in
\spectrum\ and has been summarized in a very useful way in
Appendix A of \ofer. The main results for our purposes are the
following:

\noindent $\bullet$ It might seem that the lowest-dimension non-chiral supermultiplets that can mediate SUSY breaking to the visible
sector are the $SU(2) \times SU(2)$ singlets $|A|^2$ and $|B|^2$, which have naive scaling dimension two.  It would then appear that these can
multiply visible-sector fields and break sequestering completely.  However, the
true scaling dimensions of $|A|^2, |B|^2$
grow with the 't Hooft coupling in the supergravity limit $g_s N \gg 1$, {\it{i.e.}} these operators are dual to string states in the conifold
throat.  On the other hand, the $SU(2) \times SU(2)$ non-singlet bilinears in $A, B$ are
just the harmless Noether currents discussed above.

\noindent $\bullet$ In fact, the lowest-dimension non-chiral supermultiplet
of operators invariant under $SU(2) \times SU(2)$ is given by
\eqn\lowis{ {\cal O}_{8} = W_{\alpha}^2
\overline{W}_{\dot\alpha}^2 ~,} where $W_{\alpha}$ is the standard gauge
field strength superfield whose $\theta$ expansion begins with the
gaugino $\lambda_{\alpha}$. The lowest component of ${\cal
O}_{8}$ has dimension $\Delta = 6$, while the highest component
(relevant for transmission of scalar soft masses in \leadpert) has
$\Delta = 8$. While these are the expected free-field dimensions
of the components of ${\cal O}_{8}$, it follows from the
supergravity analysis that these conformal dimensions are also
correct at strong 't Hooft coupling, though this is not guaranteed
by any known non-renormalization theorem \spectrum.
We would expect this operator to dominate the corrections to sequestering
if the SUSY-breaking VEV preserves $SU(2) \times SU(2)$.
Compared to the weak-coupling hidden sector expectation that dimension-two
bilinears multiply visible fields and mediate
 soft masses $\sim |F_{hid}|^2/M_P^2$,
${\cal O}_{8}$ will transmit soft terms that are smaller by a
factor $(\Lambda_{IR}/\Lambda_{UV})^{4}$.

\noindent $\bullet$ The lowest-dimension non-chiral supermultiplet
of operators with {\it any} $SU(2)\times SU(2)$ quantum numbers has a lowest component of
dimension $\Delta = \sqrt{28}-2 \approx 3.29$, and a highest component of dimension
$\Delta = \sqrt{28} \approx 5.29$, so we can call it ${\cal O}_{\sqrt{28}}$.
It transforms in the ${({\bf 3},{\bf 3})}$ representation of $SU(2) \times
SU(2)$.
If the SUSY-breaking VEVs spontaneously break
$SU(2) \times SU(2)$, we expect ${\cal O}_{\sqrt{28}}$ to mediate soft terms
suppressed by a factor $(\Lambda_{IR}/\Lambda_{UV})^{1.29} \ll 1$
compared to the weak-coupling
result.

Whether SUSY breaking is mediated by ${\cal O}_{8}$
or by ${\cal O}_{\sqrt{28}}$, for reasonable values of the parameters,
{\it SUSY breaking at the end of a warped throat will be effectively 
sequestered from visible sector fields living on branes in the bulk of
the Calabi-Yau.}

\newsec{Gravity-side sequestering}

The logic of conformal sequestering we have applied above is tight, but
indirect.  We will now illustrate some of the above considerations
by explicit gravity-side calculation of couplings between
separated D3-branes in the warped throat.  We demonstrate on the gravity side the suppression
of SUSY-breaking mediation to a visible-sector brane.
We also discuss the effects of compactification, analogous to
those of \S2.2.

\subsec{D3-brane couplings in a non-compact warped throat}

We will now infer the coupling between a
D3-brane at $r_{UV}$ (close to $r_{max}$) and a probe D3-brane at the tip, $r=r_0$.  To do this,
we can view the D3-brane at $r_{UV}$ as sourcing a perturbation of
the throat metric \ads, and ask how this perturbation couples to
the DBI action of the IR-brane.

In addition to the metric, we will need to know the five-form flux
$F_5$. In the unperturbed background, in the limit where we
consider the metric \ads, this is given by
\eqn\fiveform{(F_5)_{rtx^1x^2x^3} = 4 {r^3\over R^4},~~
(C_4)_{tx^1x^2x^3} = {r^4\over R^4}~} where we have chosen a gauge
for $C_4$.  The background harmonic function \eqn\harm{h(r) =
{R^4\over r^4}} is perturbed by the probe.  For values of $r \ll
r_{UV}$, it takes the form \eqn\harmtwo{h \to h + \delta h =
{R^4\over r^4} + {1\over N} {R^4\over r_{UV}^4} ~.}

We now plug this into the Born-Infeld + Chern-Simons action of the
D3-brane localized at $r_0$. Because D3-branes enjoy a no-force
condition in the pseudo-BPS backgrounds \GKP\ determined by
three-form fluxes in type IIB Calabi-Yau models, we find from
\ads, \fiveform\ that no potential is induced.  However, higher
derivative terms are indeed generated. The leading such operator
can be understood as follows.  Expanding 
\eqn\expdi{(h+\delta h)^{\pm 1/2} \approx h^{\pm 1/2}\left(1 \pm {\delta h \over 2h}\right)}
we find that the modified line element in the presence of the UV
brane is \eqn\nowis{ds^2 = h^{-1/2} \left(1 - {\delta h \over
2h}\right) \eta_{\mu\nu} dx^\mu dx^\nu + h^{1/2}\left(1 + {\delta
h \over 2h}\right) \tilde{g}_{mn} dy^m dy^n~.} 
This is a dilatation of the D3-brane worldvolume metric by a factor \eqn\dilf{\lambda = 1 -
{\delta h \over 2h}~.}

By the AdS/CFT dictionary, this perturbation of the gravitational
background should couple to a specific operator in the D3-brane
field theory.  For a D3-brane in an AdS throat, the leading term
in the Born-Infeld action that breaks conformal invariance and
hence can couple to \dilf\ is \refs{\Kabsorb,\Ferrprobe}
\eqn\klebcoup{\int d^4x \left({\delta h \over 2h}\right) \tilde{{\cal
O}} (2\pi \alpha^\prime)^2} on the worldvolume of the IR
D3-brane, with \eqn\caloei{\tilde{{\cal O}} = {2\over 3} {\rm{tr}}
\left(F_{\mu\nu} F_{\rho\nu} F_{\mu\lambda} F_{\rho\lambda} +
{1\over 2} F_{\mu\nu} F_{\rho\nu} F_{\rho\lambda} F_{\mu\lambda}
-{1\over 4} F_{\mu\nu} F_{\mu\nu} F_{\rho\lambda} F_{\rho\lambda}
-{1\over
8}F_{\mu\nu}F_{\rho\lambda}F_{\mu\nu}F_{\rho\lambda}\right) ~.}
The operator $\tilde{{\cal O}}$ is a component of the supermultiplet
of operators \eqn\tildaga{ {\cal O}_{8} = W_{\alpha}^2
\overline{W}_{\dot\alpha}^2} which, as we have already noted
\spectrum, is the lowest-dimension supermultiplet that is
invariant under $SU(2) \times SU(2)$.  The fact that \klebcoup\ is the
leading coupling, and is indeed present, was tested directly by
``scattering experiments'' off a D3-brane in \Kabsorb.

Up to the compactification effects we discuss later, this result
illustrates warped sequestering at its simplest.  The
leading operator communicating between the two D3-branes is
considerably suppressed compared to the expected quartic
cross-coupling between fields on different branes. Of course,
this does not demonstrate SUSY breaking, to which we now turn.

\subsec{SUSY breaking and sequestering in a non-compact warped throat}

A simple way of breaking SUSY in the IR of the throat is to replace the
IR D3-brane above with  an anti-D3-brane. The Born-Infeld
logic goes through unchanged, and the leading cross-coupling is
still mediated by an operator of the form \tildaga.  However, in
this case we can borrow direct results for the probe D3/anti-D3
potential in a warped throat, as derived in \KKLMMT. The result,
in the absence of a stabilizing potential for the UV D3-brane, is
an attractive potential \eqn\potis{ V(r_{UV}) = 2 T_3
\left({r_0\over{R}}\right)^4 \left(1-{1\over N}{r_0^4\over
r_{UV}^4}\right) \, .}
This is in agreement with our expectation that the
leading operator transmitting SUSY breaking is ${\cal O}_{8}$.
In the presence of warping, even if the cross-coupling involved a
hidden sector operator whose lowest component had dimension two (the
free-field value for a chiral-antichiral bilinear),
we would expect suppression of the interaction by four powers of
the IR warp-factor $r_0/R$, because this factor already suppresses the hidden
SUSY-breaking vacuum energy density (the leading term of \potis) due to warping as in \RSone.

In \potis, we instead see a suppression of cross-couplings by
{\it eight} powers of the IR warp factor.
This indicates that
the leading hidden operator that communicates with the UV-brane
fields has dimension eight (or its lowest component has dimension
six). This is completely consistent with our AdS/CFT reasoning,
and with the results of \Ferrprobe.

The fact that ${\cal O}_8$ and not ${\cal O}_{\sqrt{28}}$ has appeared here can be traced to our consideration of the {\it{radial}} potential.
The radial position of a D3-brane corresponds to a scalar field that is invariant under the $SU(2)\times SU(2)$ global symmetry, 
so an operator coupling pairs of such fields, for a separated brane-antibrane pair, will also be invariant; and ${\cal O}_{8}$ is the lowest-dimension $SU(2)\times SU(2)$-invariant non-chiral operator.  We expect that in more general circumstances
(and perhaps after including leading $\alpha^\prime$ corrections to the
tree-level gravity solution),
couplings mediated by ${\cal O}_{\sqrt{28}}$ will be present.

Note that the visible brane fluctuations, corresponding to fluctuations of the UV brane position, $r_{UV}$, not only get
highly-suppressed
SUSY breaking masses by Taylor expanding \potis\ in $\delta r_{UV}$ to quadratic
order, but also a suppressed
tadpole term linear in $\delta r_{UV}$. This reflects
an instability of the visible D3-brane: the D3-brane is attracted toward the anti-D3-brane,
so in principle we do not have a legitimate ground state. This unwanted
feature can be cured by more realistic visible sector model-building in the
Calabi-Yau bulk, and in particular by stabilization of the branes on which the
visible sector is localized.  Here we are only interested in illustrating the
high degree of suppression of visible-sector SUSY-breaking masses.

In the earlier work \KKLMMT, the focus was on using the
inter-brane force \potis\ to drive inflation.  The conclusion was
that although the cross-couplings \potis\ are small, the full
potential for $r_{UV}$ receives additional contributions and is
generically much steeper than \potis.  This could be attributed to
a conformal coupling of the D3-brane modes to the four-dimensional
curvature, which is nontrivial during inflation.
A crucial difference is that for sequestering, one is interested
in the inter-brane interactions in a background with approximately
vanishing vacuum energy. Therefore, after tuning to solve the
cosmological constant problem (for example, by an appropriate
shift of the flux superpotential), the conformal coupling becomes a
negligible effect.

\subsec{Compactification effects}

As in the cases of low warping in \S2, we expect that
compactification will result in mixing between brane fields and bulk
moduli, and that generically this mixing will lead to further cross-couplings
between separated brane stacks (or brane and antibrane stacks). However,
locality now implies that such induced couplings must still proceed via
throat modes to communicate with IR-branes or antibranes. The most
important modes are still the ones we discussed above, and therefore
we still expect the same degree of sequestering, although the numerical
coefficients may differ.  Most throat modes lead to strong
sequestering of cross-couplings between UV-branes and IR-branes, and the dominant
effects will arise through exchange of the Kaluza-Klein gauge bosons
corresponding to the $SU(2) \times SU(2)$ isometries of the throat,
which have vanishing $AdS_5$ mass.  These are dual to the conserved
currents discussed in \S3.2,
and thus they mediate cross-couplings
in an unsuppressed fashion.  However, these effects are harmless, since by the result of \SS\
they do not mediate IR SUSY breaking.

It is instructive to test these results directly on the gravity side.  In a general warped compactification, the full K\"ahler potential can be rather complicated, and explicit formulae have proved elusive (see \WS\ for a recent
discussion).
However, there is a very simple system that is rich enough to manifest the desired phenomenon, the emergence of unsuppressed
but nevertheless harmless cross-couplings after compactification, and for which we know the relevant formulae ${\it exactly}$.
This system is the ${\cal N}=4$ supersymmetric toroidal orientifold with a distribution of D3-branes discussed in \Anisimov\ and in the introduction.
If we separate the D3-branes into two stacks, one to provide the visible
sector and the remainder to form the hidden sector, the gravitational backreaction of the latter stack forms a warped throat,
as noted originally in \Herman.  This throat is dual to the ${\cal N}=4$ SYM dynamics of the hidden stack.
But despite the strong warping, the high degree of supersymmetry protects the form of the K\"ahler potential \torus, and it is unaltered by
the continuation to large $g_sN$. As noted in the introduction, this system
does have formally unsequestered cross-couplings.  But, these are precisely couplings of the visible sector to the conserved currents of
the hidden sector, associated to the throat isometries, and in the presence of hidden SUSY breaking these couplings pose no threat (at least in
the circumstances described in \SS).  The fact that even in the unwarped regime the
${\cal N}=4$ orientifold enjoys this milder form of sequestering is ``explained'' by
the fact that it {\it must} do so in the warped regime $g_s N \gg 1$ (by our previous
arguments), and
the non-renormalization
of the K\"ahler potential then forces it to do so even in the unwarped limit.
One can also directly see that the
cross-couplings vanish in the non-compact limit (in which the radii are taken to infinity), which agrees with the observation of \S2.1 that the D3-branes do not see each other prior to compactification.

\newsec{Classes of warped sequestering}

In our discussion of warped sequestering, we have focused on concrete examples arising
in compactifications that incorporate the warped deformed conifold \KS.  However, while this is a
useful and common example, we wish to emphasize that there is by now a wide spectrum of other
throats (understood in different levels of detail) that could be more useful for various model
building purposes.

The two main ingredients that go into specifying a model, independent of visible-sector model building, are:
(a) the description of an appropriate AdS/CFT dual pair that governs the throat dynamics
(describes the approximate fixed point controlling the RG cascade), and (b) the incorporation of an
appropriate SUSY-breaking sector in the IR region of the throat.
Both are subjects of intense research, characterized by rapid recent progress.

In considering possibilities for (a), it is useful to keep in mind that the main dangers to
sequestering involve global currents in the CFT.  Any ${\cal N}=1$ CFT has a $U(1)_R$ symmetry,
which does not pose any danger to sequestering. But one could wish to find examples that
lack additional continuous
global symmetries, because when such symmetries are present we must take extra care to ensure that
deformations that break hidden sector conformal invariance are also symmetric \SS\ (as
we checked in the case of the conifold throat).
The smaller the global symmetry group, the fewer such checks
we need to perform.
  Several new classes of Sasaki-Einstein solutions to string theory have been
discovered recently, including the $Y^{p,q}$ and $L^{p,q,r}$ series \refs{\ypq,\lpqr}.
Similarly, cascades based on diverse solutions have been described in {\it{e.g.}} \refs{\ypqcasc,\dpcasc,
\bertolini,\Doran}.
While the relevant CFTs in these new classes
are characterized by less symmetry than the conifold,
they still possess continuous global symmetries beyond the minimal $U(1)_R$.

However, it is possible to exhibit simple, infinite families of dual pairs that only
possess $U(1)_R$.  Even ${\cal N}=1$ orbifolds of the ${\cal N}=4$ SYM theory \refs{
\DM,\orbCFT,\LNV}
by non-Abelian discrete subgroups of $SU(3)$ are useful in this regard.
Indeed, the
series of discrete groups $\Delta_{3n^2}$ and $\Delta_{6n^2}$ give rise to orbifold CFTs
with only the $U(1)_R$ subgroup of the $SO(6)$ symmetry of ${\cal N}=4$ SYM surviving
(the relevant quiver gauge theories were derived in detail in \GR).
One can see this as follows.  The group $\Delta_{3n^2}$, for instance, is generated by
\eqn\genone{\alpha: (z_1,z_2,z_3) \to (e^{2\pi i/n} z_1, e^{-2\pi i/n} z_2, z_3)~,}
\eqn\gentwo{\beta: (z_1,z_2,z_3) \to (z_1, e^{2\pi i/n} z_2, e^{-2\pi i/n} z_3)~,}
and
\eqn\genthree{\gamma: (z_1,z_2,z_3) \to (z_3,z_1,z_2)~.}
The maximal abelian subgroup of $SO(6)$ is generated by independent phase rotations of
the $z_i$.  It is easy to see that $\gamma$ only commutes with the single $U(1)$ that
rotates the $z_i$ by equal phases.  Therefore, these nonabelian orbifolds preserve only
a single $U(1)_R$ subgroup of the $SO(6)$ symmetry of ${\cal N}=4$ SYM.
It
would be interesting to make cascading solutions controlled by these CFTs but ending smoothly
in the IR \students.  These series in some sense provide a counterexample to the expectation in
\DFG\ that typical CFTs will contain additional continuous symmetries, with the corresponding
dangerous supermultiplets of currents.

Supersymmetry breaking at the end of RG cascades (or at the end of warped throats, in the dual
gravity language) has also been a focus of recent research.  In the warped deformed conifold,
the models of \KPV\ provide examples.  The quivers that characterize simple $\Z_k$ quotients
of the conifold were described in \Uranga, and cascades governed by the associated CFTs
together with SUSY breaking mechanisms at the end of the throat have been investigated in
\ABFK.  Other cascades with SUSY breaking, associated with cones
over del Pezzo surfaces, have been examined in
\refs{\bertolini,\berenstein,\Franco}.  Many of these models suffer from a runaway vacuum
\IS; for the state of the art in obtaining stable vacua
this way, see \refs{\Argurio,\diaconescu,\wijnholt}.

Finally, although the bulk of the literature on warped compactifications in string theory
focuses on type IIB vacua, this is only for reasons of convenience: the solutions are conformally
Calabi-Yau and the most familiar examples of AdS/CFT arise in this context.
We expect that further research will uncover similar rich families of warped compactifications
in the type IIA, heterotic, and eleven-dimensional limits, which could be equally or more promising
for the construction of fully realistic models.

\newsec{Conclusion}

We have argued that warped sequestering is robustly attainable in
string theory, and that sequestered interactions between D3-branes
are possible even in some unwarped models. Warped throats are a
rather generic feature of type IIB flux compactifications
\refs{\KS,\GKP,\DD,\GKT,\Hebecker}, so we conclude that one should
be able to design, without undue difficulty, concrete string theory
models of SUSY breaking in which the flavor problem is vitiated by sequestering.
Such models merit further detailed exploration.

One of our motivations for pursuing this work
was the advent of mirage mediation \mirage, a scenario where anomaly mediation
is combined with moduli mediation in a warped flux compactification.  We hope to
have provided a firm foundation for discussion of the circumstances under
which such models can lead to flavor-blind supersymmetry breaking.
We note that the sequestering due to warped throats is insensitive to the details
of the standard model construction in the bulk of the Calabi-Yau space.
In particular, while D7-brane fields are not sequestered from D3-brane fields in
the non-compact, unwarped limit, SUSY breaking at the end of a throat can be sequestered from D7-branes
in a suitably warped type IIB flux compactification.  This makes it seem likely that 
relatively realistic models of warped sequestering incorporating unification of coupling constants can be constructed
in string theory.

\bigskip\
\centerline{\bf{Acknowledgements}}
\smallskip\

S.K. is grateful to O. Aharony, Y. Antebi,
K. Choi, M. Dine, S. Giddings, D. Martelli, M. Mulligan, N. Seiberg, and J. Sparks
for useful discussions.  He happily acknowledges the hospitality of KITP Santa
Barbara, where this
work was initiated.  The research of S.K. was supported by a David and Lucile Packard Foundation
Fellowship, by the National Science Foundation under grant number 0244728, and by the Department of Energy under contract
DE-AC02-76SF00515.
L.M. thanks D. Baumann, J. Maldacena, and A. Murugan for
very helpful discussions on related subjects, and is grateful to KITP Santa Barbara and the theory groups at Stanford University,
the University of Texas, and the University of Michigan for their hospitality.  The research of L.M. was supported in part by the Department of Energy under grant
DE-FG02-90ER40542.  R.S. is
grateful for discussions with M. Schmaltz on
related topics.  The research of R.S. was supported by the National Science Foundation
grant NSF-PHY-0401513 and by the Johns Hopkins Theoretical Interdisciplinary
Physics and Astrophysics Center.

\listrefs
\end